\documentclass[twocolumn,showpacs,preprintnumbers,amsmath,amssymb,aps,prb]{revtex4}
\usepackage{graphicx}
\begin{document}

\title{
Collective Transport Properties of Driven Skyrmions with Random Disorder
} 
\author{
C. Reichhardt, D. Ray, and C. J. Olson Reichhardt 
} 
\affiliation{
Theoretical Division and Center for Nonlinear Studies,
Los Alamos National Laboratory, Los Alamos, New Mexico 87545, USA\\ 
} 

\date{\today}
\begin{abstract}
We use particle-based simulations to 
examine the static and driven collective phases of skyrmions
interacting with random quenched disorder.
We show that non-dissipative effects due to the Magnus
term reduce the depinning threshold and strongly affect the skyrmion
motion and the nature of the dynamic phases.  
The quenched disorder 
causes the Hall angle to become drive-dependent in the moving skyrmion
phase, while different flow regimes produce distinct signatures
in the transport curves.  
For weak disorder, the skyrmions form a pinned crystal and depin elastically,
while for strong disorder the system forms a pinned amorphous state 
that depins plastically. 
At high drives the skyrmions can dynamically reorder into a moving crystal, 
with the onset of reordering determined by
the strength of the Magnus term.
\end{abstract}
\maketitle

There are a wide variety of systems that can be effectively modeled as
collectively interacting particles moving over quenched disorder, 
where there is a transition from a pinned state
to a sliding state
under an applied drive. 
Examples include driven
incommensurate charge density waves \cite{1,2}, Wigner crystals
\cite{3,4}, colloids driven over various types of substrates
\cite{5,6,7}, and vortices in type-II superconductors
\cite{8,9,10,11,12}.  In many of these systems the particle-particle
interactions are repulsive, so that in the absence of disorder 
a hexagonal crystal of particles forms.  When quenched disorder is present, the
particles may retain hexagonal or mostly hexagonal order if the disorder is 
weak,
while stronger disorder can lead to a proliferation of
topological defects creating an amorphous or glassy state
\cite{8,9,13}.  
If an additional external
driving force $F_{D}$ is now applied to particles 
in a pinned state, 
there is a critical drive value 
$F_{c}$ 
known as the depinning threshold 
above which the particles begin to move.
For weak disorder
the particles generally depin elastically and retain their original
neighbors \cite{1,2,5,7,9},
but for strong disorder the depinning 
is often  
plastic with particles continuously changing neighbors over time,  
forming
a fluctuating liquid-like state \cite{4,5,6,9,10,11,12}.
Additionally, 
if the depinning occurs from a disordered pinned state, 
there can be 
dynamic 
structural transitions at drives well above $F_{c}$, 
where the particles
can dynamically order into a moving anisotropic crystal or moving
smectic phase \cite{9,10,11,14,15,16,17,18,19,20}.  
Such transitions occur because 
the effectiveness of the pinning is reduced in the 
drive direction 
when the particles are strongly driven \cite{10}.  
Transitions from pinned to plastic and from plastic to
dynamically ordered states have been observed experimentally through
features in transport measures \cite{1,6,9,11}, 
neutron scattering \cite{14},
changes in noise fluctuations \cite{11}, 
and direct imaging experiments \cite{18}.

Recently a new type of system, 
skyrmions in chiral magnets, has been realized that
can be effectively 
characterized 
as particle-like objects interacting with random disorder
\cite{21,22,23}. 
Neutron scattering experiments showed evidence for 
triangular skyrmion lattices in bulk MnSi samples \cite{22}, 
and subsequent Lorentz microscopy experiments \cite{22,23,24} 
produced direct observations of
skyrmion lattices.  
Imaging measurements indicate that 
as an externally applied magnetic field is increased, 
a low-density, partially-disordered skyrmion phase appears 
out of a helical phase, 
and is subsequently transformed
into a denser ordered triangular lattice of skyrmions;  
at high fields, the skyrmion number decreases 
and the system eventually 
enters a uniform ferromagnetic state \cite{22,23}.  
In addition to their particle-like nature,
another similarity that skyrmions have to vortices in
type-II superconductors is that they 
can be externally
driven by the application of a current \cite{25}. Experimental
transport measurements have shown \cite{26} that it is possible to
obtain skyrmion velocity vs applied driving force curves and that
there is a finite depinning threshold.  
The current-driven motion of skyrmions has also been directly
experimentally imaged \cite{27,28}.  

One aspect of skyrmions that makes them very distinct
from other collectively driven systems in random disorder is the
pronounced non-dissipative component in
the skyrmion equation of motion 
arising from the Magnus term \cite{23,29,30,31}. 
This term 
causes skyrmions to move in a
direction {\it perpendicular} to the applied driving force,
and it has been
argued to be the cause of the relatively low depinning threshold
observed for skyrmions \cite{23,29,30,31}. 
The strength of the Magnus term 
is denoted by
$\beta$ while that of the damping is denoted by $\alpha$. 
In superconducting vortex systems,
$\beta/\alpha \ll 1.0$ so the non-dissipative terms are very weak
\cite{8}; 
in contrast, skyrmion systems typically 
have $\beta/\alpha$ values
ranging from $10$ to $40$ so that the Magnus term dominates the 
skyrmion dynamics. 
This 
makes skyrmions a unique system in which to explore non-dissipative collective
dynamics.  
Beyond basic science issues,
skyrmions may be useful for a range of applications \cite{32} 
which will require an
understanding of skyrmion dynamics in the presence of random
disorder.

Here we utilize a recently developed particle-based model for skyrmion
dynamics to study collective skyrmion behaviors in the presence
of random quenched disorder for varied disorder strengths and varied 
$\beta/\alpha$ values.
We show that the Magnus term 
reduces the depinning threshold and 
induces orbits with a circular character 
for skyrmions moving in and across pinning sites.  
We also show
that collective skyrmion-skyrmion interactions play an important role
in the Magnus-induced reduction of the depinning threshold.
We find that the Magnus term introduces
a Hall angle for the skyrmion motion relative to the direction of the
external drive, and that 
the addition of quenched disorder strongly reduces this Hall
angle, particularly just above depinning. 
At higher drives,
the velocity-force curves 
both parallel and perpendicular to the drive show
distinctive features 
associated with
a transition into a dynamically
ordered state.
We also find disorder-induced
transitions from a skyrmion glass to a skyrmion crystal phase, 
observable
as a function of disorder strength or skyrmion density.

\begin{figure}
\includegraphics[width=\columnwidth]{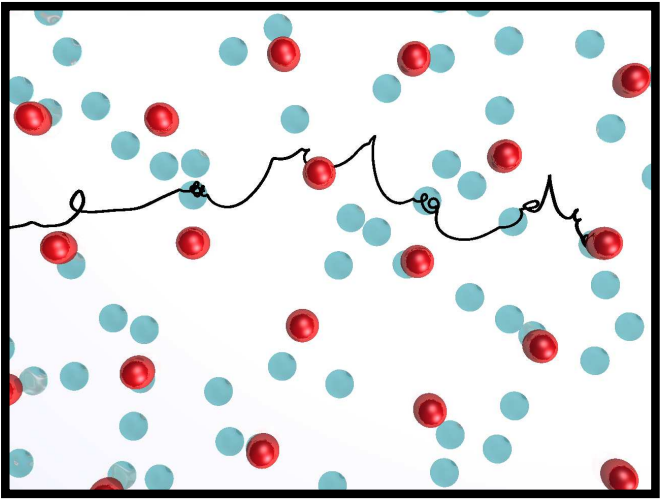}
\caption{ 
Real-space image of skyrmions (dark red dots) driven through 
randomly arranged pinning sites (light blue dots).
Here $F_{p} = 0.03$ and $F_{D} = 0.0125$; 
the system is in a moving plastic flow phase. 
The trajectory of a single moving skyrmion is highlighted,  
showing 
spiraling type motions inside the pinning sites.  
}
\label{fig:1}
\end{figure}

{\it Simulation and System---} 
We simulate skyrmions interacting with random disorder using a recently
developed particle model \cite{31}, where the dynamics of a single
skyrmion $i$ is governed by the following equation:
\begin{equation}  
\alpha\frac{d {\bf R}_{i}}{dt} = 
{\bf F}^{M}_{i} +  {\bf F}^{ss}_{i} + {\bf F}^{sp}_{i} + {\bf F}^{D}_{i}  .
\end{equation} 
Here $\frac{d {\bf R}_{i}}{dt} = {\bf v}_{i}$ is the skyrmion
velocity. 
The damping term $\alpha$ 
arises from the damping of 
the spin precession and damping of electrons localized in the
skyrmions.  
The Magnus term ${\bf F}^{M}_{i} = \beta {\bf v}_{i} \times
{\hat z}$ 
produces a force oriented perpendicular to the skyrmion velocity. 
We impose the constraint $\alpha^2 + \beta^2 = 1$  
in order to maintain a constant magnitude of the skyrmion velocity  
for varied ratios of $\beta/\alpha$. 
For systems 
such as MnSi, $\beta/\alpha \approx 10$;\cite{31} 
in this work, unless otherwise noted, we take 
$\beta/\alpha=9.962$, corresponding to 
Magnus-dominated particle dynamics. 
The skyrmion-skyrmion interaction force is 
${\bf F}^{ss}_{i} = \sum^{N_{s}}_{j=1} \hat{\bf r}_{ij} K_{1}(R_{ij})$ 
where $R_{ij}=|{\bf r}_i - {\bf r}_j|$, 
$\hat{\bf r}_{ij}=({\bf r}_i - {\bf r}_j)/R_{ij}$,  
and $K_{1}$ is the modified Bessel function
which falls off exponentially for large $R_{ij}$.
The pinning force ${\bf F}^{sp}_{i}$ arises from 
randomly placed, non-overlapping 
harmonic 
traps of size $R_{p}=0.3$ with a maximum pinning force of $F_{p}$. 
The driving term
${\bf F}^{D}$ 
represents a Lorentz force from an externally applied current bias 
interacting with the 
emergent magnetic flux carried by the skyrmions \cite{26}. 

Our system is of size $L \times L$ with $L=36$, 
has periodic boundary conditions in the $x$ and $y$ directions, 
and contains $N_s$ skyrmions and $N_p$ pinning sites.   
The pin density $\rho_p=N_p/L^2$ is fixed at 0.3, 
and the skyrmion density $\rho_s=N_s/L^2$ equals 0.1 
unless otherwise noted.
We apply a slowly increasing driving force ${\bf F}^{D}$ 
and measure the average skyrmion
velocity in the direction parallel (perpendicular) to the applied drive, 
$\langle V_{\rm drive} \rangle$  
($\langle V_{\perp} \rangle$). 
The Hall angle is calculated from these velocities 
as $\theta = \tan^{-1}R$, 
where $R = \langle V_{\perp}\rangle/\langle V_{\rm drive}\rangle$. 
For a single skyrmion driven in the absence of disorder, 
$R =\beta/\alpha$, 
so that $\theta=84.25^{\circ}$ for our simulations 
in the clean limit;  
in contrast, a vortex with $\beta \approx 0.0$ 
has $\theta=0$ and moves along the drive direction. 

{\it Results and Discussion---} 
In Fig.~1 we show a snapshot of 
skyrmions moving through random disorder in a small portion of a system with
$F_{p} = 0.03$, and $F_{D} = 0.01$. 
Since this drive is well above the depinning threshold $F_c=0.02$,  
all the skyrmions are in motion; 
however, we highlight the trajectory of a single skyrmion, 
which shows that the skyrmion 
undergoes a swirling or circular motion within each pinning site 
it encounters before escaping from the pin.
The circular motion arises due to the Magnus term: 
the force from each pinning site points 
toward its center, 
but under Magnus-dominated dynamics this force is largely
perpendicular to the skyrmion velocity, 
causing the skyrmions to circle around the inner edges of the pinning
sites.
In the overdamped limit with $\beta/\alpha \ll 1$, 
a skyrmion entering a pinning site 
would quickly travel to the bottom of the potential well and be strongly
pinned. 
We find that the escape of a skyrmion from a pinning site is strongly
affected by motion excited via interactions with the surrounding skyrmions.

\begin{figure}
\includegraphics[width=\columnwidth]{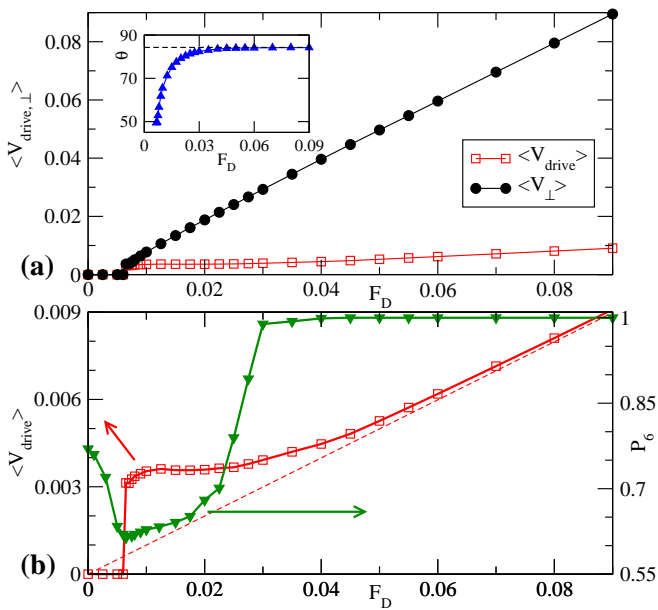}
\caption{ 
(a) Average skyrmion velocity in the drive direction,  
$\langle V_{\rm drive} \rangle$ (open squares),  
and perpendicular to the drive, 
$\langle V_{\perp} \rangle$ (filled circles),  
vs $F_{D}$ for system with $F_{p} = 0.03$.
Inset: The Hall angle $\theta$ vs $F_{D}$; 
the dotted line is the result for the clean system.  
(b) $\langle V_{\rm drive}\rangle$ (open squares) and 
the fraction of sixfold coordinated particles $P_6$
(filled triangles) vs 
$F_{D}$ for the same system as in (a); 
the dashed line shows $\langle V_{\rm drive}\rangle$ for the clean system. 
There is a dynamical ordering transition 
into a moving crystal state at $F_D\approx 0.03$.
}
\label{fig:2}
\end{figure}

We first examine how the Hall angle is affected by the 
presence of random disorder. 
To characterize this, in Fig.~2(a) we plot 
$\langle V_{\rm drive} \rangle$ and $\langle V_{\perp} \rangle$
versus $F_{D}$ for a system with $F_{p} = 0.03$. 
For $F_{D} < 0.00625$, the skyrmions are pinned in a disordered 
arrangement.  
Just above depinning, 
$\langle V_{\rm drive} \rangle \approx \langle V_{\perp} \rangle$, 
so that $R \approx 1.0$ is much less than the clean-limit 
value.
In general we find that adding quenched
disorder decreases $\langle V_{\perp} \rangle$ 
and increases $\langle V_{\rm drive} \rangle$ compared to the clean limit,
indicating that
the average skyrmion flow is rotated back toward the 
drive direction.
This is illustrated in the inset of Fig.~2(a), which 
shows that the Hall angle is well below the clean-limit 
value at low drives and approaches it at higher drives. 
The origin of this behavior is a ratcheting effect experienced by 
the skyrmions as they encounter pinning sites.
When a drive is applied, the equilibrium pinned position of a skyrmion 
is displaced from the pin's center in the direction of the drive,
and a skyrmion driven over a pin will be pulled toward this equilibrium point.
Consequently, on average, the outgoing trajectory of a skyrmion 
emerging from a pin is offset in the drive direction relative to
its incoming trajectory. 
When $F_{D} \lesssim F_{p}$, a skyrmion can spend 
a lot of time interacting with a pinning site, enhancing the size of this
offset,
while for higher drives the offset is diminished. 
 
At the drive value shown in Fig.~1, the skyrmions form a disordered state; 
however, at higher drives they can dynamically order into a moving crystal.  
In Fig.~2(b) we plot 
the fraction of six-fold
coordinated particles $P_{6}$ as a function of $F_{D}$ 
along with $\langle V_{\rm drive} \rangle$ 
for the same system as in Fig.~2(a).
$P_{6}$ is smallest right at depinning where the particles are most
disordered, but rapidly approaches 1 
around $F_{D} \approx 0.025$ 
indicating dynamical reordering of the system. 
Simultaneously, $\langle V_{\rm drive} \rangle$ 
acquires a linear dependence on $F_{D}$,
with values approximating the clean-limit value. 

\begin{figure}
\includegraphics[width=\columnwidth]{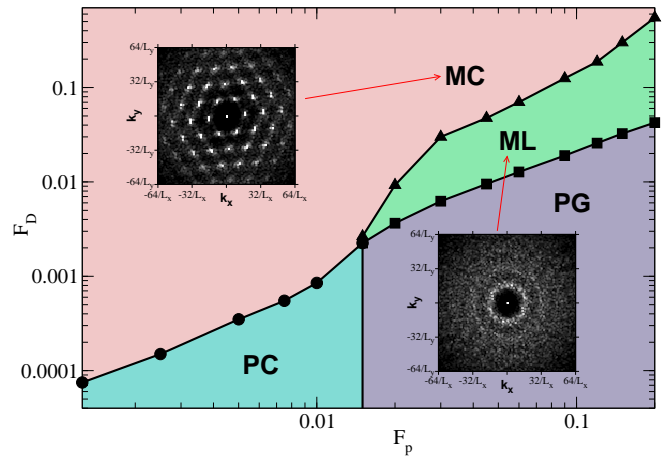}
\caption{ 
Dynamical phase diagram for $F_D$ vs $F_p$
highlighting the different skyrmion phases.
PC: pinned crystal;
PG: pinned amorphous glass;
ML: moving liquid;
MC: moving crystal.
Circles: elastic depinning from PC to MC;
squares: plastic depinning from PG to ML; 
triangles: dynamical ordering transition from ML to MC.
Upper inset: structure factor $S(k)$ of the skyrmion positions 
in the MC state.
Lower inset: $S(k)$ in the ML state.
}
\label{fig:3}
\end{figure}

In Fig.~3 we plot a dynamical phase diagram highlighting the static and
dynamic phases for skyrmions 
as $F_{D}$ and $F_{p}$ are varied. 
For $F_{p} < 0.015$ the skyrmions form a pinned triangular
crystal (PC) which depins elastically 
to a moving crystal state (MC) as the drive increases.
For $F_{p} > 0.015$ at low drives, we instead find an amorphous 
pinned skyrmion glass (PG) which depins {\it plastically} 
with increasing drive 
into a fluctuating moving skyrmion liquid (ML).
The lower inset of Fig.~3 shows that the structure factor $S(k)$ of the 
ML phase 
has a liquid-like ring.
As $F_{D}$ increases further, 
the moving liquid transitions into a moving skyrmion crystal state
with sixfold ordering, as shown in the plot of $S(k)$ in the upper
inset of Fig.~3. 
This phase diagram has similarities to that found for a driven
vortex system \cite{19}; 
however, the skyrmions reorder into a moving crystal
rather than a moving smectic state. 
The vortex moving smectic state 
forms when the pinning remains effective in the direction
transverse to the vortex motion while being weakened in the
direction of motion, subjecting the vortices to
an effective anisotropic temperature.
In the skyrmion case, 
additional fluctuations induced by the Magnus 
force 
reduce the
transverse pinning in the moving state, 
giving a more isotropic effective temperature and allowing the skyrmions to
form a more
isotropic moving structure.  
We expect that neutron scattering experiments could be used to observe the 
transition from a skyrmion glass to a moving skyrmion crystal state 
as a function of external drive.

\begin{figure}
\includegraphics[width=\columnwidth]{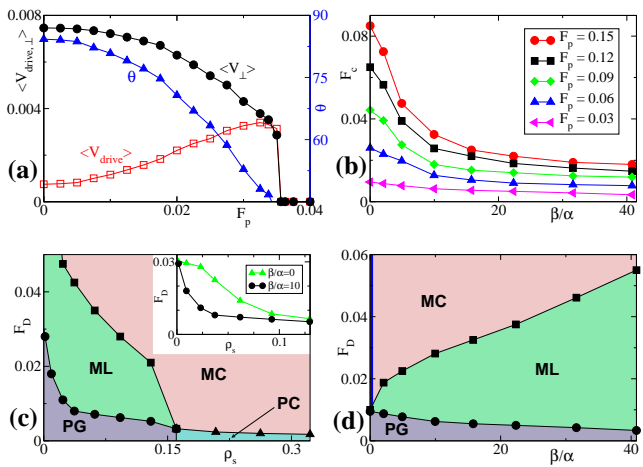}
\caption{
(a) $\langle V_{\rm drive}\rangle$, $\langle V_{\perp}\rangle$, and $\theta$
as a function of disorder strength $F_{p}$, 
at fixed drive $F_{D} = 0.0075$. 
(b) The critical depinning force $F_{c}$ vs $\beta/\alpha$ 
for $F_{p} = 0.15$, 0.12, 0.09, 0.06, and 0.03, from top to bottom.
(c) Phase diagram for $F_{D}$ vs $\rho_s$
at 
$F_{p} = 0.03$, 
highlighting the transition from PG to ML (circles),
ML to MC (squares), and 
PC to MC (triangles). 
Inset: The depinning threshold (PG-ML transition line)
for Magnus-dominated (circles) 
and damping-dominated (triangles) particle dynamics.
(d) Phase diagram of $F_{D}$ vs $\beta/\alpha$ for 
$F_{p} = 0.03$, 
showing transitions from PG to ML (circles) and
ML to MC (squares). 
Blue line along upper left axis indicates moving smectic state 
for overdamped particles. 
}
\label{fig:4}
\end{figure}

To better understand how the pinning affects 
skyrmion motion, 
in Fig.~4(a) we plot $\langle V_{\rm drive}\rangle$ and 
$\langle V_{\perp}\rangle$ for fixed 
$F_{D} = 0.0075$ 
and varied pinning strength. 
As $F_{p}$ increases 
and the skyrmions transition from MC to ML to PG
(as shown in Fig.~3), 
$\langle V_{\rm drive} \rangle$ increases and 
$\langle V_{\perp} \rangle$ decreases
from the clean-limit values 
so that $\langle V_{\rm drive} \rangle$ actually slightly exceeds 
$\langle V_{\perp} \rangle$ at $F_{p} = 0.035$, just before the 
skyrmions become pinned.
The Hall angle accordingly decreases as $F_{p}$ is increased. 
In Fig.~4(b) we plot the critical
depinning force $F_{c}$ vs $\beta/\alpha$ for varied $F_{p}$.  
As the dynamics become 
increasingly Magnus-dominated
for higher $\beta/\alpha$, $F_{c}$ monotonically decreases, 
confirming that inclusion of Magnus forces lowers the depinning
threshold. 
This effect is more prominent for higher values of $F_{p}$. 

In Fig.~4(c) we plot the phase diagram for $F_{D}$ versus 
skyrmion density $\rho_s$
at fixed 
$F_{p} = 0.03$. 
At low skyrmion densities, 
a disordered pinned state forms that 
depins plastically into a moving liquid
and then orders into a moving crystal at higher drives; 
at higher $\rho_s$, we find a pinned crystal state  
that depins elastically directly into a moving crystal state.  
Notably, the depinning threshold
is very close to $F_{p}$ 
at the lowest simulated density (near the single
skyrmion limit),  
but falls off rapidly as $\rho_s$ increases.
This 
indicates
that {\it collective} skyrmion-skyrmion interactions 
play a crucial role in producing a low depinning threshold. 
This is further emphasized in
the inset of
Fig.~4(c) where we show how the plastic depinning line 
is altered when we go from Magnus-dominated to damping-dominated 
particle dynamics. 
At low $\rho_s$, $F_c$ is nearly the same for the two systems; however,
once the skyrmion-skyrmion interactions become important for higher
$\rho_s$, the 
falloff in the depinning threshold with increasing $\rho_s$ 
is much more rapid 
in the Magnus-dominated system.
In Fig.~4(d) we show how the dynamical phase diagram 
changes as the dynamics become increasingly Magnus-dominated by 
plotting the phases as a function of $F_{D}$ and $\beta/\alpha$.
When $\beta/\alpha = 0$, the behavior is the same as that of a
vortex system and the system reorders into
a moving smectic rather than a moving crystal. 
As $\beta/\alpha$ increases, the depinning threshold drops
while the drive at which dynamical reordering occurs increases
due to the enhanced swirling motion of the
skyrmions in the liquid state, 
which produces an effective temperature that is more isotropic but also
larger in magnitude compared to the overdamped case.

{\it Summary---} 
We have investigated the depinning dynamics of skyrmions
interacting with random disorder, utilizing a recently developed
particle-based skyrmion model. We find that the 
Magnus-dominated dynamics typical of skyrmions decreases the
depinning threshold due to the swirling orbits followed by skyrmions
interacting with pinning sites.
Skyrmion-skyrmion scattering 
tends to 
excite such orbits 
and thus also plays an important role in reducing the
depinning threshold.  
For increasing disorder 
strength we find
transitions
from a pinned skyrmion crystal to an amorphous skyrmion glass.
We also show that the Hall angle deviates from its clean-limit value
for strong pinning or 
weak driving.
At high drives,
in contrast to superconducting vortices which dynamically reorder into a 
smectic state, the skyrmions undergo a dynamical phase
transition from a fluctuating driven liquid to a moving crystal. 
This occurs because the Magnus term tends to make the fluctuations, 
and the resulting effective temperature, experienced
by the skyrmions more isotropic with respect to the direction of the
external drive compared to a damping-dominated system.
Features in the transport response such as velocity-force curves or
structure factor measurements can be used to identify signatures of
the different dynamical phases.

\acknowledgments
We thank S.-Z. Lin for useful discussions.
This work was carried out under the auspices of the 
NNSA of the 
U.S. DoE
at 
LANL
under Contract No.
DE-AC52-06NA25396 and through the LANL/LDRD program.

\end{document}